\documentclass[12pt,a4paper]{article}

\usepackage{longtable}
\usepackage{ifthen} 
\newboolean{pdflatex}
\setboolean{pdflatex}{false} 

\newboolean{articletitles}
\setboolean{articletitles}{true} 

\newboolean{uprightparticles}
\setboolean{uprightparticles}{false} 

\newcommand{\cleoc}{{C}{L}{E}{O}{-}{c}}

\newcommand{\prt}[1]{\ensuremath{#1}} 
\newcommand{\DDbar}{\prt{D\overline{D}}}
\newcommand{\fourpi}{\prt{\pi^+\pi^-\pi^+\pi^-}}

\newcommand{\secref}[1]{Sec.~\ref{#1}}

\usepackage{microtype}
\usepackage{lineno}  
\usepackage{xspace} 
\usepackage{caption} 

\usepackage{graphicx}  
\usepackage{color}
\usepackage{colortbl}
\graphicspath{{./figs/*}} 

\usepackage{amsmath} 
\usepackage{amssymb}
\usepackage{amsfonts}
\usepackage{upgreek} 

\newcommand*\patchAmsMathEnvironmentForLineno[1]{%
\expandafter\let\csname old#1\expandafter\endcsname\csname #1\endcsname
\expandafter\let\csname oldend#1\expandafter\endcsname\csname
end#1\endcsname
 \renewenvironment{#1}%
   {\linenomath\csname old#1\endcsname}%
   {\csname oldend#1\endcsname\endlinenomath}%
}
\newcommand*\patchBothAmsMathEnvironmentsForLineno[1]{%
  \patchAmsMathEnvironmentForLineno{#1}%
  \patchAmsMathEnvironmentForLineno{#1*}%
}
\AtBeginDocument{%
\patchBothAmsMathEnvironmentsForLineno{equation}%
\patchBothAmsMathEnvironmentsForLineno{align}%
\patchBothAmsMathEnvironmentsForLineno{flalign}%
\patchBothAmsMathEnvironmentsForLineno{alignat}%
\patchBothAmsMathEnvironmentsForLineno{gather}%
\patchBothAmsMathEnvironmentsForLineno{multline}%
\patchBothAmsMathEnvironmentsForLineno{eqnarray}%
}

\usepackage{hyperref}    
\usepackage[all]{hypcap} 

\usepackage{xspace} 
\usepackage{upgreek}

\def\MagUp {\mbox{\em Mag\kern -0.05em Up}\xspace}

\ifthenelse{\boolean{uprightparticles}}%
{

 \def\Ppi         {\ensuremath{\uppi}\xspace}

 \def\PDelta      {\ensuremath{\Delta}\xspace}                 
 \def\PXi      {\ensuremath{\Xi}\xspace}                 
 \def\PLambda      {\ensuremath{\Lambda}\xspace}                 
 \def\PSigma      {\ensuremath{\Sigma}\xspace}                 
 \def\POmega      {\ensuremath{\Omega}\xspace}                 
 \def\PUpsilon      {\ensuremath{\Upsilon}\xspace}                 
 
 \def\PB      {\ensuremath{\mathrm{B}}\xspace}                 
                  
 \def\PD      {\ensuremath{\mathrm{D}}\xspace}

 \def\PK      {\ensuremath{\mathrm{K}}\xspace}

 \def\Pe      {\ensuremath{\mathrm{e}}\xspace}

 \def\Pi      {\ensuremath{\mathrm{i}}\xspace}

}
{

 \def\Ppi         {\ensuremath{\pi}\xspace}

 \mathchardef\PDelta="7101
 \mathchardef\PXi="7104
 \mathchardef\PLambda="7103
 \mathchardef\PSigma="7106
 \mathchardef\POmega="710A
 \mathchardef\PUpsilon="7107
                  
 \def\PB      {\ensuremath{B}\xspace}                 
                  
 \def\PD      {\ensuremath{D}\xspace}

 \def\PK      {\ensuremath{K}\xspace}

 \def\Pe      {\ensuremath{e}\xspace}

 \def\Pi      {\ensuremath{i}\xspace}

}

\makeatletter
\ifcase \@ptsize \relax
  \newcommand{\miniscule}{\@setfontsize\miniscule{4}{5}}
\or
  \newcommand{\miniscule}{\@setfontsize\miniscule{5}{6}}
\or
  \newcommand{\miniscule}{\@setfontsize\miniscule{5}{6}}
\fi
\makeatother

\DeclareRobustCommand{\optbar}[1]{\shortstack{{\miniscule (\rule[.5ex]{1.25em}{.18mm})}
  \\ [-.7ex] $#1$}}

\def\epem       {{\ensuremath{\Pe^+\Pe^-}}\xspace}

\def\pion   {{\ensuremath{\Ppi}}\xspace}

\def\pip    {{\ensuremath{\pion^+}}\xspace}
\def\pim    {{\ensuremath{\pion^-}}\xspace}

\def\kaon    {{\ensuremath{\PK}}\xspace}
  \def\Kbar    {{\kern 0.2em\overline{\kern -0.2em \PK}{}}\xspace}

\def\KorKbar    {\kern 0.18em\optbar{\kern -0.18em K}{}\xspace}

\def\Km      {{\ensuremath{\kaon^-}}\xspace}

  \def\Dbar    {{\kern 0.2em\overline{\kern -0.2em \PD}{}}\xspace}
\def\D       {{\ensuremath{\PD}}\xspace}

\def\DorDbar    {\kern 0.18em\optbar{\kern -0.18em D}{}\xspace}
\def\Dz      {{\ensuremath{\D^0}}\xspace}
\def\Dzb     {{\ensuremath{\Dbar{}^0}}\xspace}
\def\Dp      {{\ensuremath{\D^+}}\xspace}
\def\Dm      {{\ensuremath{\D^-}}\xspace}

\def\B       {{\ensuremath{\PB}}\xspace}
\def\Bbar    {{\ensuremath{\kern 0.18em\overline{\kern -0.18em \PB}{}}}\xspace}

\def\BorBbar    {\kern 0.18em\optbar{\kern -0.18em B}{}\xspace}

\def\Bub     {{\ensuremath{\B^-}}\xspace}

\def\Bm      {{\ensuremath{\Bub}}\xspace}

  \def\Y#1S{\ensuremath{\PUpsilon{(#1S)}}\xspace}

\def\Lbar        {{\ensuremath{\kern 0.1em\overline{\kern -0.1em\PLambda}}}\xspace}
\def\LorLbar    {\kern 0.18em\optbar{\kern -0.18em \PLambda}{}\xspace}

\def\to                 {\ensuremath{\rightarrow}\xspace}

\def\CP                {{\ensuremath{C\!P}}\xspace}

\def\AT#1     {\ensuremath{A_{\mathrm{T}}^{#1}}\xspace}           

\def\C#1      {\ensuremath{\mathcal{C}_{#1}}\xspace}                       
\def\Cp#1     {\ensuremath{\mathcal{C}_{#1}^{'}}\xspace}                    
\def\Ceff#1   {\ensuremath{\mathcal{C}_{#1}^{\mathrm{(eff)}}}\xspace}        
\def\Cpeff#1  {\ensuremath{\mathcal{C}_{#1}^{'\mathrm{(eff)}}}\xspace}       
\def\Ope#1    {\ensuremath{\mathcal{O}_{#1}}\xspace}                       
\def\Opep#1   {\ensuremath{\mathcal{O}_{#1}^{'}}\xspace}                    

\newcommand{\tev}{\ifthenelse{\boolean{inbibliography}}{\ensuremath{~T\kern -0.05em eV}\xspace}{\ensuremath{\mathrm{\,Te\kern -0.1em V}}}\xspace}
\newcommand{\gev}{\ensuremath{\mathrm{\,Ge\kern -0.1em V}}\xspace}
\newcommand{\mev}{\ensuremath{\mathrm{\,Me\kern -0.1em V}}\xspace}
\newcommand{\kev}{\ensuremath{\mathrm{\,ke\kern -0.1em V}}\xspace}
\newcommand{\ev}{\ensuremath{\mathrm{\,e\kern -0.1em V}}\xspace}
\newcommand{\gevc}{\ensuremath{{\mathrm{\,Ge\kern -0.1em V\!/}c}}\xspace}
\newcommand{\mevc}{\ensuremath{{\mathrm{\,Me\kern -0.1em V\!/}c}}\xspace}
\newcommand{\gevcc}{\ensuremath{{\mathrm{\,Ge\kern -0.1em V\!/}c^2}}\xspace}
\newcommand{\gevgevcccc}{\ensuremath{{\mathrm{\,Ge\kern -0.1em V^2\!/}c^4}}\xspace}
\newcommand{\mevcc}{\ensuremath{{\mathrm{\,Me\kern -0.1em V\!/}c^2}}\xspace}

\def\pb {\ensuremath{\mathrm{ \,pb}}\xspace}
\def\invpb {\ensuremath{\mbox{\,pb}^{-1}}\xspace}

\newcommand{\stat}{\ensuremath{\mathrm{\,(stat)}}\xspace}

\newcommand{\chisq}{\ensuremath{\chi^2}\xspace}

\def\gsim{{~\raise.15em\hbox{$>$}\kern-.85em
          \lower.35em\hbox{$\sim$}~}\xspace}
\def\lsim{{~\raise.15em\hbox{$<$}\kern-.85em
          \lower.35em\hbox{$\sim$}~}\xspace}

\def\tell1  {TELL1\xspace}
\def\ukl1   {UKL1\xspace}

\newcommand{\ie}{\mbox{\itshape i.e.}\xspace}

\usepackage{mciteplus}

\usepackage{pifont}
\definecolor{darkgreen}{rgb}{0.0,0.5,0.0}
\definecolor{darkred}{rgb}{0.5,0.0,0.0}

\usepackage[labelfont=bf]{caption}

\usepackage{booktabs}
\usepackage[nottoc]{tocbibind}

\begin{document}

\renewcommand{\thefootnote}{\fnsymbol{footnote}}
\setcounter{footnote}{1}


\begin{titlepage}

\vspace*{-1.5cm}

\begin{tabular*}{\linewidth}{lc@{\extracolsep{\fill}}r}
 \\
 & &  \\  
 & & \today \\ 
 & & \\
\hline
\end{tabular*}

\vspace*{4.0cm}

{\bf\boldmath\huge
\begin{center}
  Amplitude analysis of $D^{0} \to \fourpi \, $ decays using CLEO-c data
\end{center}
}

\vspace*{2.0cm}

\begin{center}
P.~d'Argent\footnote{Speaker}$^1$,
J.~Benton$^2$,
J.~Dalseno$^2$,
E.~Gersabeck$^1$,
S.T.~Harnew$^2$,
P.~Naik$^2$, \\
C.~Prouve$^2$, 
J.~Rademacker$^2$,
N.~Skidmore$^2$
\bigskip\\
{\it\footnotesize
$ ^1$Physikalisches Institut, Ruprecht-Karls-Universit\"{a}t Heidelberg, Heidelberg, Germany\\
$ ^2$H.H. Wills Physics Laboratory, University of Bristol, Bristol, United Kingdom\\
}
\end{center}

\vspace{\fill}

\begin{abstract}
  \noindent
The resonant substructure of the decay $\Dz \to \fourpi$ is studied by performing a full five-dimensional amplitude analysis. 
Preliminary results based on data collected by the CLEO-c detector are presented.
This is the largest dataset of $\Dz \to \fourpi$ decays analysed in this way to-date. 
The two most significant contributions are $\Dz \to a_{1}(1260)^{+} \, \pim $ and $\Dz \to \rho(770)^{0} \, \rho(770)^{0}$. The line shape, mass and width of the $a_{1}(1260)$ resonance are determined, and model-independent studies of the line shapes of several resonant contributions are preformed. 
\end{abstract}

\vspace*{2.0cm}

\begin{center}
  Presented at \\ 
  VIII International Workshop On Charm Physics \\
   Bologna, Italy, 5-9 September, 2016
\end{center}

\vspace{\fill}

\end{titlepage}

\pagestyle{empty}  


%

\renewcommand{\thefootnote}{\arabic{footnote}}
\setcounter{footnote}{0}

\pagestyle{plain} 
\setcounter{page}{1}
\pagenumbering{arabic}


\section{Introduction}
\label{sec:introduction}

We present preliminary results of the amplitude analysis of the decay \prt{\Dz\ \to \fourpi}.
This is an independent analysis using the CLEO-c legacy dataset. 
The decay mode \prt{\Dz\ \to \fourpi} has the potential to make an important contribution to the determination of
the \CP-violating phase $\gamma/\phi_3 \equiv - \arg(V_{ud}V^*_{ub}/V_{cd}V^*_{cb})$ in \prt{\Bm \to \D \Km}
decays~\cite{DalitzGamma1, Rademacker}. 
The all-charged final state (impossible in three-body decays of \Dz) particularly
suits the environment of hadron collider experiments, such as LHCb. 
The sensitivity to the weak phase can be significantly improved with a 
measured amplitude
model, either to be used directly in the $\gamma$ extraction, or in
order to optimise model-independent
approaches~\cite{DalitzGamma1,Atwood:coherenceFactor,Bondar:2005ki,coherenceFromMixing2}. 
A study of the rich resonance structure of this four-body mode is also of considerable interest in its own right
providing valuable insights into strong interactions at low energies.


\section{Event Selection}
\label{sec:selection}

The data set consists of $\epem$ collisions produced by the Cornell Electron Storage Ring (CESR) at $\sqrt s \approx 3.77 \gev$ corresponding to an integrated luminosity of $818 \invpb$ and collected with the CLEO-c detector. 
At \cleoc, \D mesons are created in the process \prt{e^+ e^- \to \psi(3770) \to \DDbar}, where $\DDbar=\Dz\Dzb$ or \Dp\Dm. 
We select events where one neutral \D meson decays into four pions.
Signal selection is performed by using the standard CLEO-c selection criteria as described in Ref.~\cite{PhysRevD.76.112001} on the candidate tracks.
We reject candidates consistent with a $\Dz \to K_{S}^{0} (\to \pip \pim) \, \pip \, \pim$ decay by requiring
$\vert m(\pip \pim) - m_{K_{S}^{0}} \vert > 7.5 \mev$ for any $\pip \, \pim$ combination.
The flavour of the initial \D mesons (\Dz or \Dzb) is determined by identifying individual charged kaons from the accompanying \D decay.
Assuming these kaons are the result of Cabbibo-favoured \D meson decays, 
the flavour of both \D mesons can be inferred with a mistag probability of $\omega = (4.5 \pm 0.5) \%$ \cite{KKpipi}.
The number of signal events that pass the selection is $7\,536 \pm 74$.


\section{Amplitude analysis}
\label{sec:fit}

An amplitude analysis is performed in oder to isolate the various intermediate states contributing to the decay $D^{0} \to \pip \pim \pim \pip$.
We use the isobar approach which 
assumes that a multi-body process can be factorized into subsequent quasi-two-body decays \cite{isobar1}.
In this model, the intermediate state amplitudes can be parameterized as a product of
form factors, $B_{L}$, included for each vertex of the decay tree,
Breit-Wigner propagators, $ T_{R}$,  included for each resonance, $R$,
and an overall angular distribution represented by a spin factor, $S$,
\begin{equation}
	\mathcal A_{i}=  B_{L_{D}} \, [B_{L_{R_{1}}}  \,  T_{R_{1}}] \, [B_{L_{R_{2}}} \,  T_{R_{2}}]  \,   S_{i}  \, .
	\label{eq:amp4}
\end{equation}
For $B_{L}$ we use Blatt-Weisskopf damping factors \cite{Bl}
which depend on the relative orbital angular momentum, $L$, among the daughter particles.
The spin factors are constructed 
in a covariant tensor formalism \cite{Zemach,Filippini,Zou}.
The total amplitude for the $D^{0} \to \pip \pim \pim \pip$ decay is given by the coherent sum over all 
intermediate state amplitudes weighted by the complex coefficients $a_{i}$ 
to be measured from data:
\begin{equation}
	\mathcal A_{D^{0}} = \sum_{i}  a_{i} \, \mathcal A_{i}  \, .
	\label{eq:ampBar}
\end{equation}
Similarly, the amplitude for $\overline{D^{0}}$ decays is given by $\mathcal A_{\overline{D^{0}}} = \sum_{i} \overline{a_{i}} \, \overline{\mathcal A_{i}}$.
In our default fit, we assume that there is no \CP violation in this decay channel such that the 
amplitude coefficients for $D$ and $\overline{D}$ decays are identical ($a_{i} = \overline{a_{i}}$).
This assumption is tested in \secref{sec:CPV}.
As the \fourpi final state involves two pairs of indistinguishable pions, the amplitudes are Bose-symmetrized.

\section{Signal model selection}
\label{sec:LASSO}

\begin{figure}[ht!]
	\centering
	\includegraphics[width=0.7\textwidth, height = 6cm]{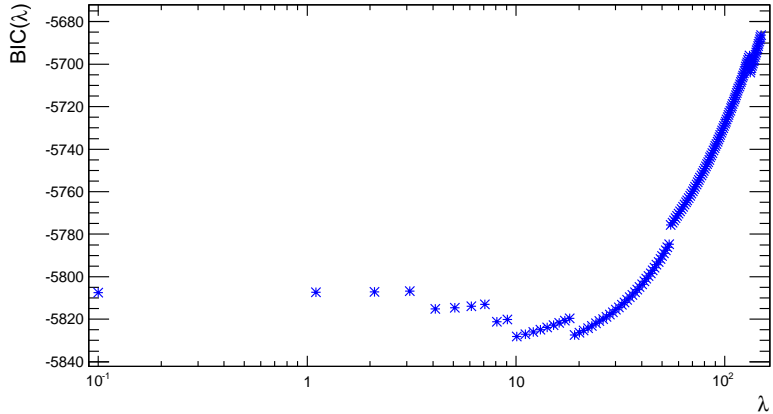} 
	\caption{The BIC value as function of the LASSO parameter $\lambda$.}
	\label{fig:BIC}
\end{figure}

The large amount of possible sub-processes in a four-body decay necessitates a model-building procedure in order to select the most significant contributions.
start with a large pool of amplitudes and use the
Least Absolute Shrinkage and Selection Operator (LASSO \cite{Tibshirani94regressionshrinkage}) approach to limit the model complexity,
as proposed in Ref.~\cite{Guegan:2015mea}. 
In this method, the likelihood function is extended by a penalty term
\begin{equation}
	-2 \, \log \mathcal L + \lambda \, \sum_{i} \sqrt{ \int \vert a_{i} \, \mathcal A_{i} \vert^{2} \, \text{d}\Phi_{4}  },
\end{equation}
where the integral is over the phase-space of the decay. 
The LASSO term
shrinks the amplitude coefficients
towards zero.
The amount of shrinkage is controlled by the parameter $\lambda$ to be tuned on data.
Higher values for $\lambda$ encourage sparse models, \ie models with only a few non-zero amplitude coefficients.
The optimal value for $\lambda$ is found by minimizing the Bayesian information criteria (BIC \cite{BIC}),
\begin{equation}
	\text{BIC}(\lambda) = - 2 \, \log \mathcal L + r  \, \log N_{\rm sig},
\end{equation}
where $N_{\rm sig}$ is the number of signal events and $r$ is the number of amplitudes with a decay fraction above 
a certain threshold. 
In this way, the optimal $\lambda$ balances
the fit quality ($- 2 \, \log  \mathcal L$) against the model complexity.
Figure \ref{fig:BIC} shows the distribution of BIC values obtained by scanning over $\lambda$
where we choose the threshold to be $0.5 \%$.
The set of amplitudes selected using the optimal value for $\lambda$, which is found to be $\lambda = 20$, 
is henceforth called the LASSO model. 

\clearpage

\section{Results}
\label{sec:results}

\begin{figure}[hp]
	\includegraphics[width=0.49\textwidth, height = 5cm]{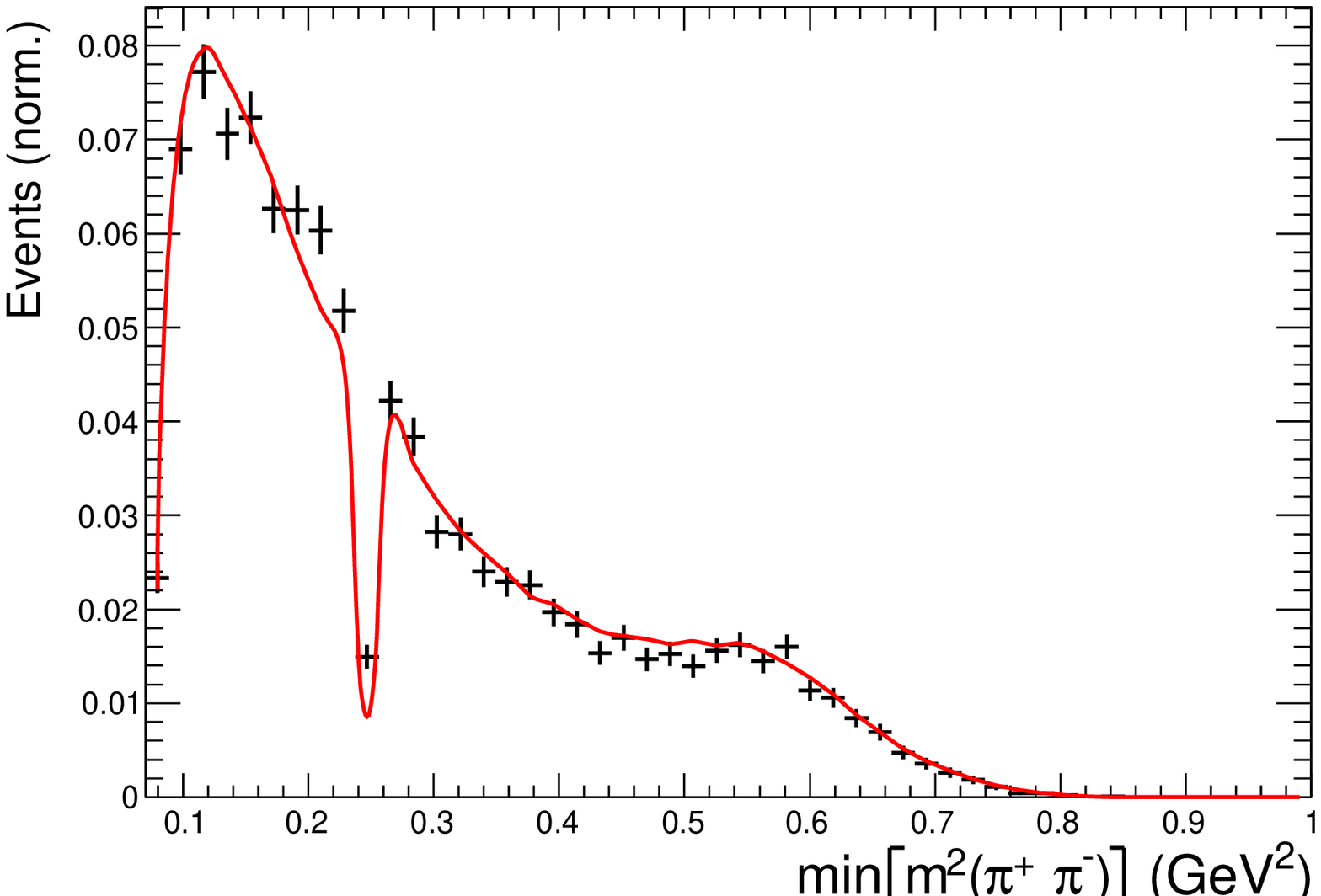} 
	\includegraphics[width=0.49\textwidth, height = 5cm]{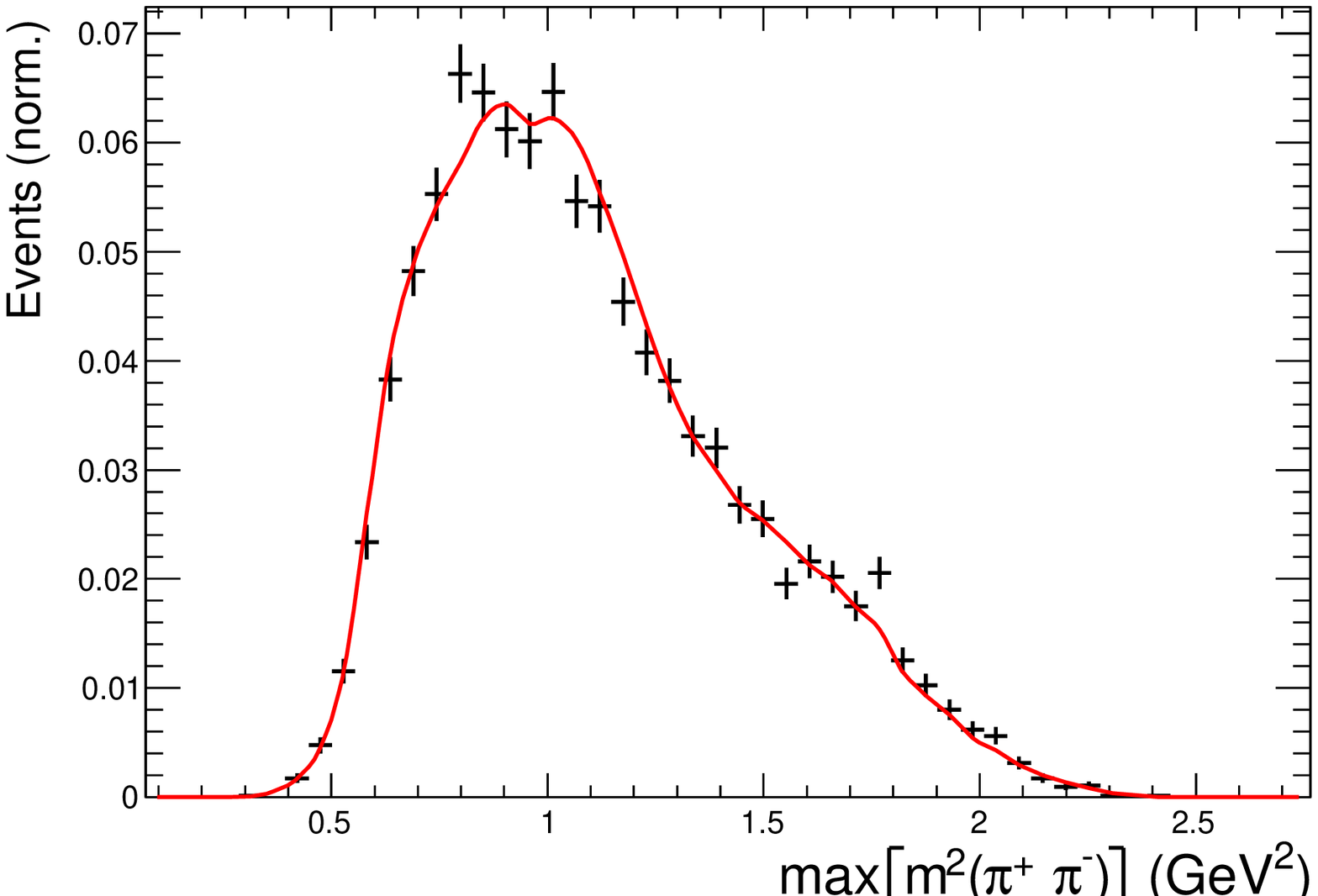} 

	\includegraphics[width=0.49\textwidth, height = 5cm]{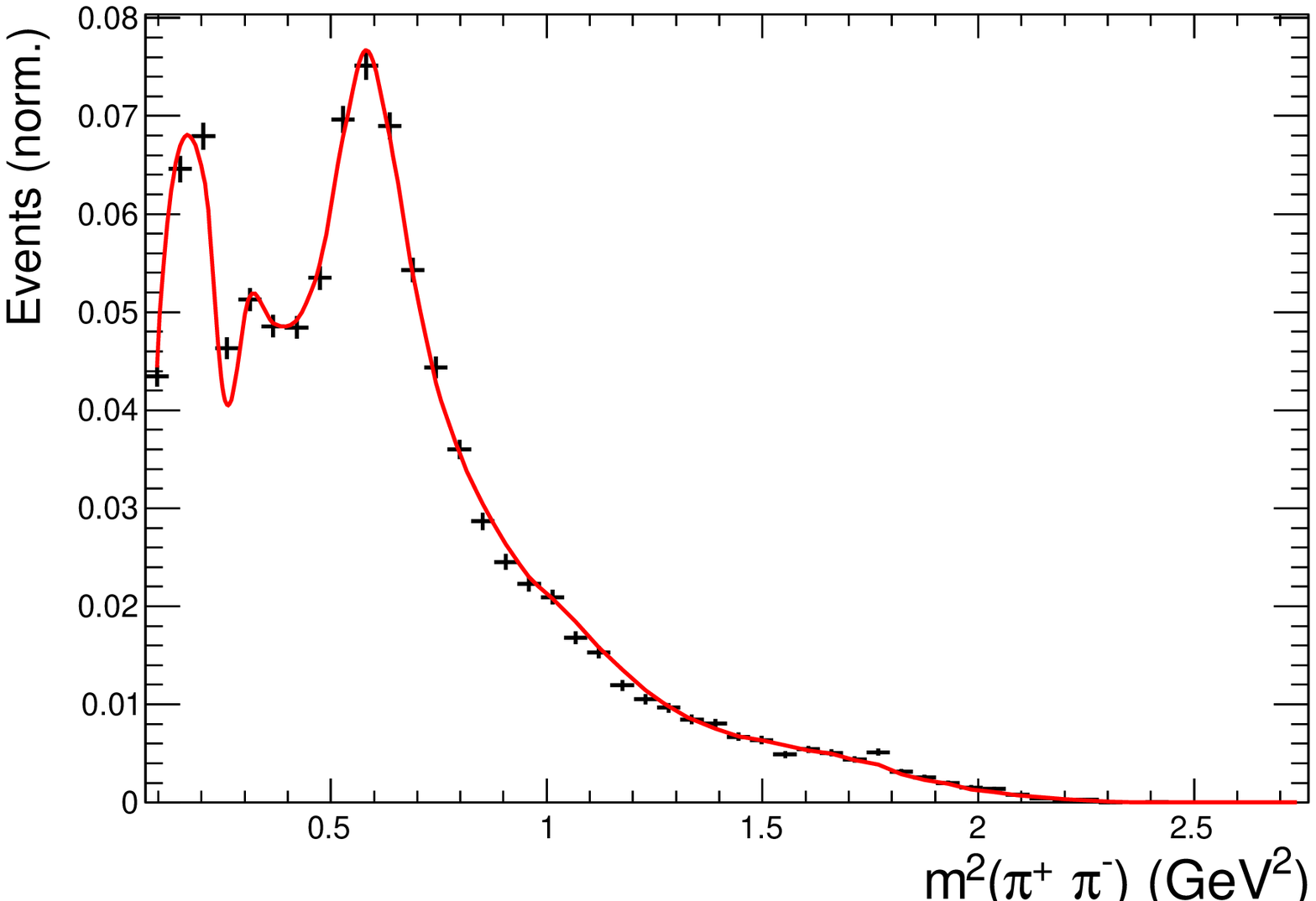} 
	\includegraphics[width=0.49\textwidth, height = 5cm]{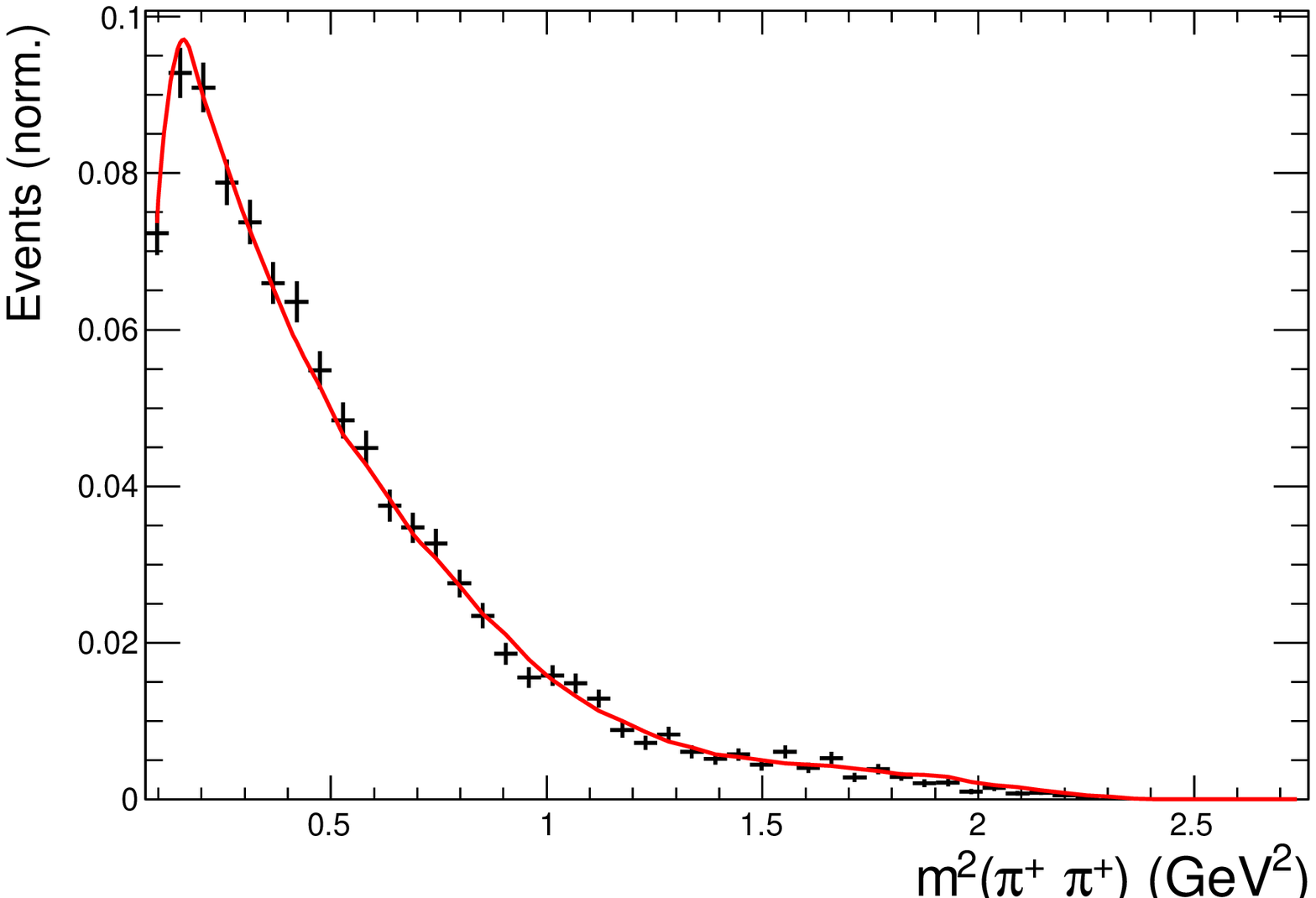} 

	\includegraphics[width=0.49\textwidth, height = 5cm]{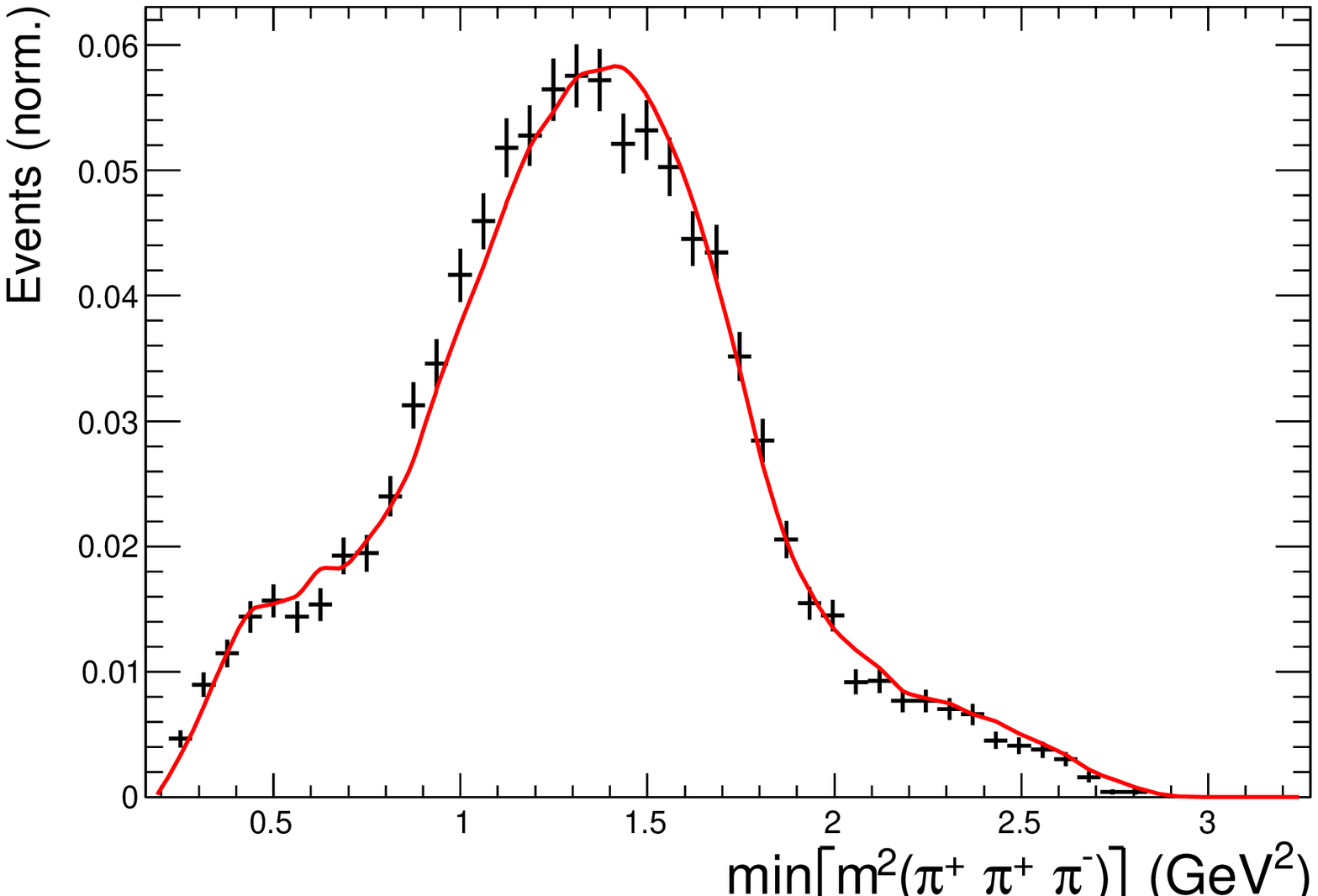} 
	\includegraphics[width=0.49\textwidth, height = 5cm]{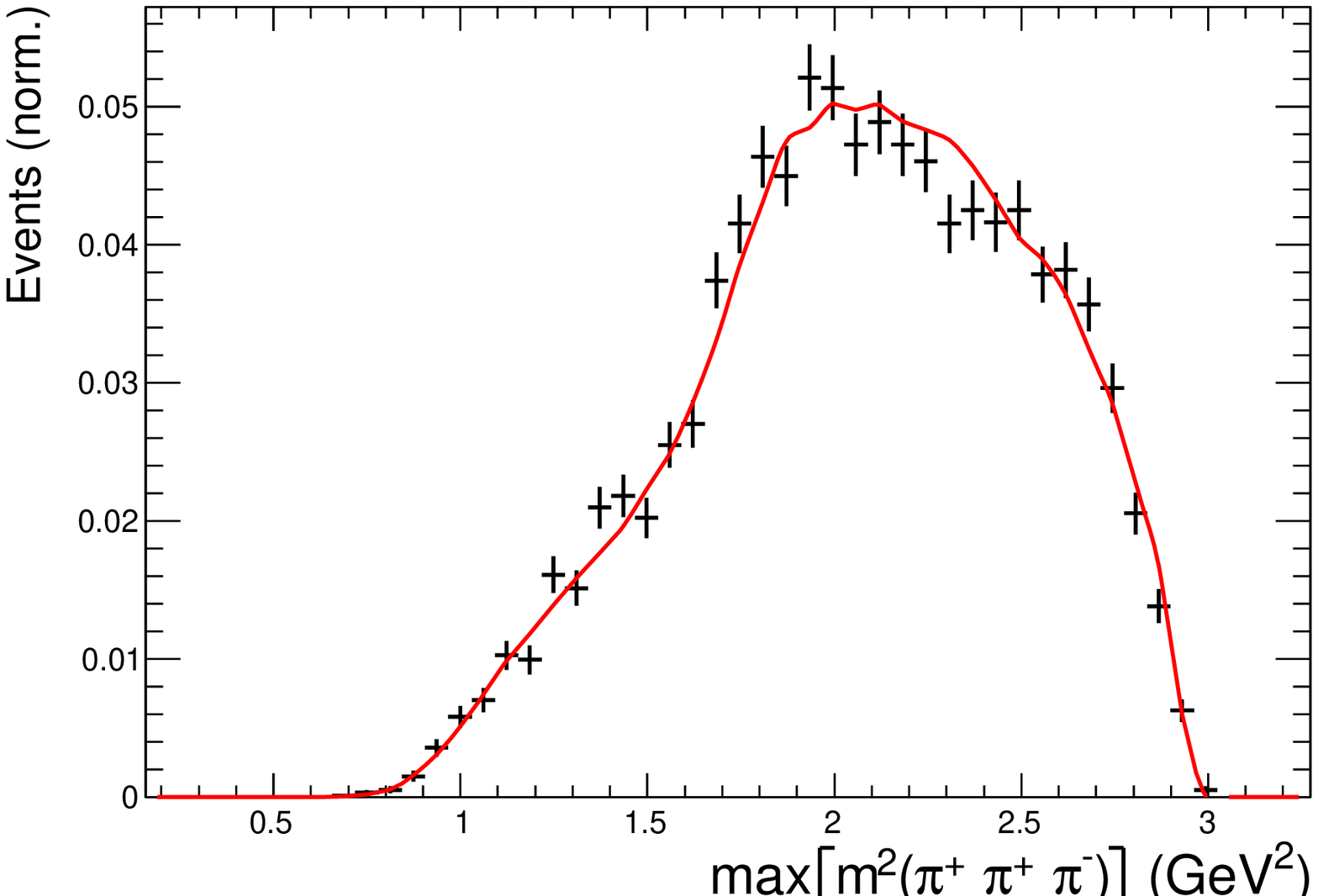} 
	
	\includegraphics[width=0.49\textwidth, height = 5cm]{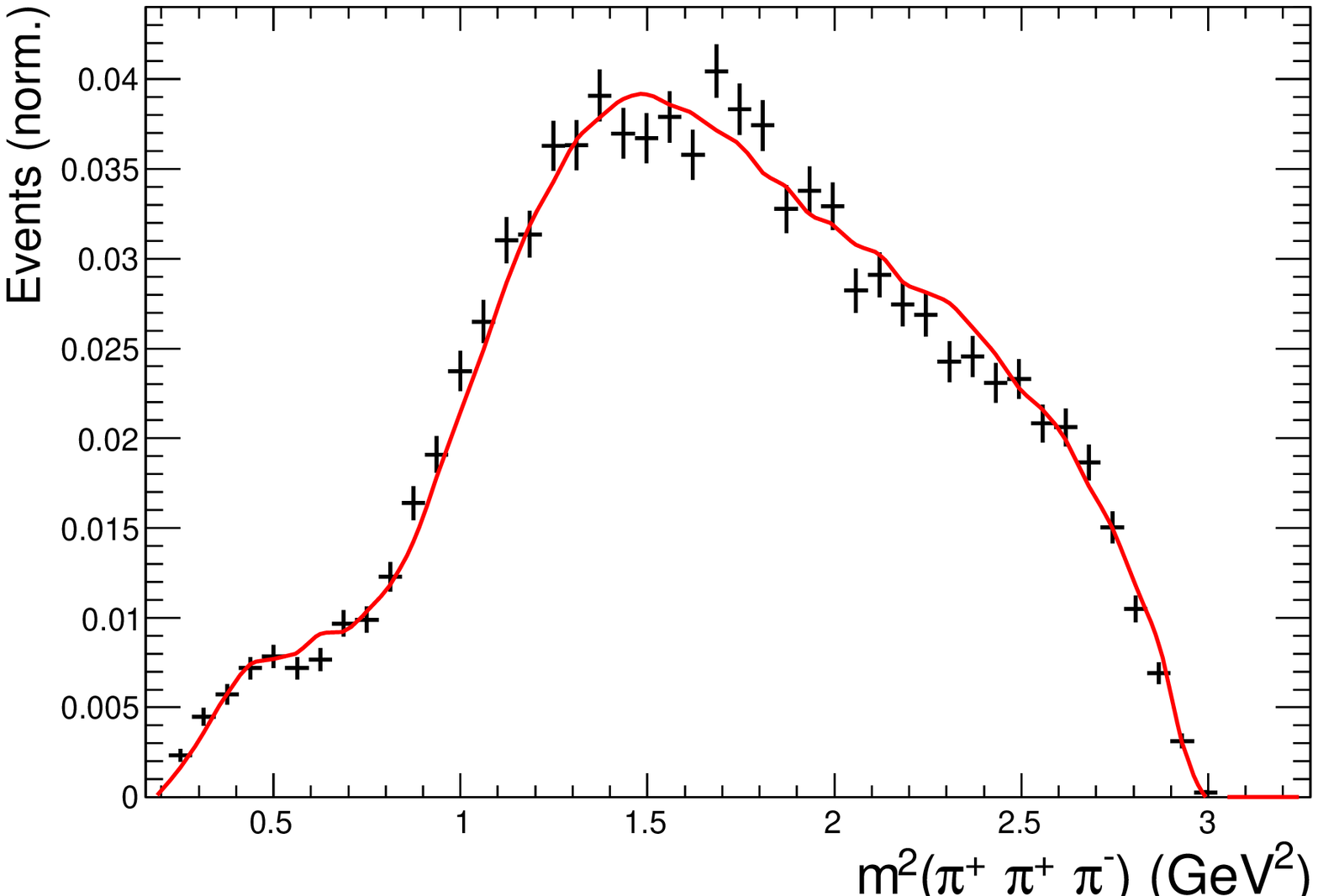} 
	\includegraphics[width=0.49\textwidth, height = 5cm]{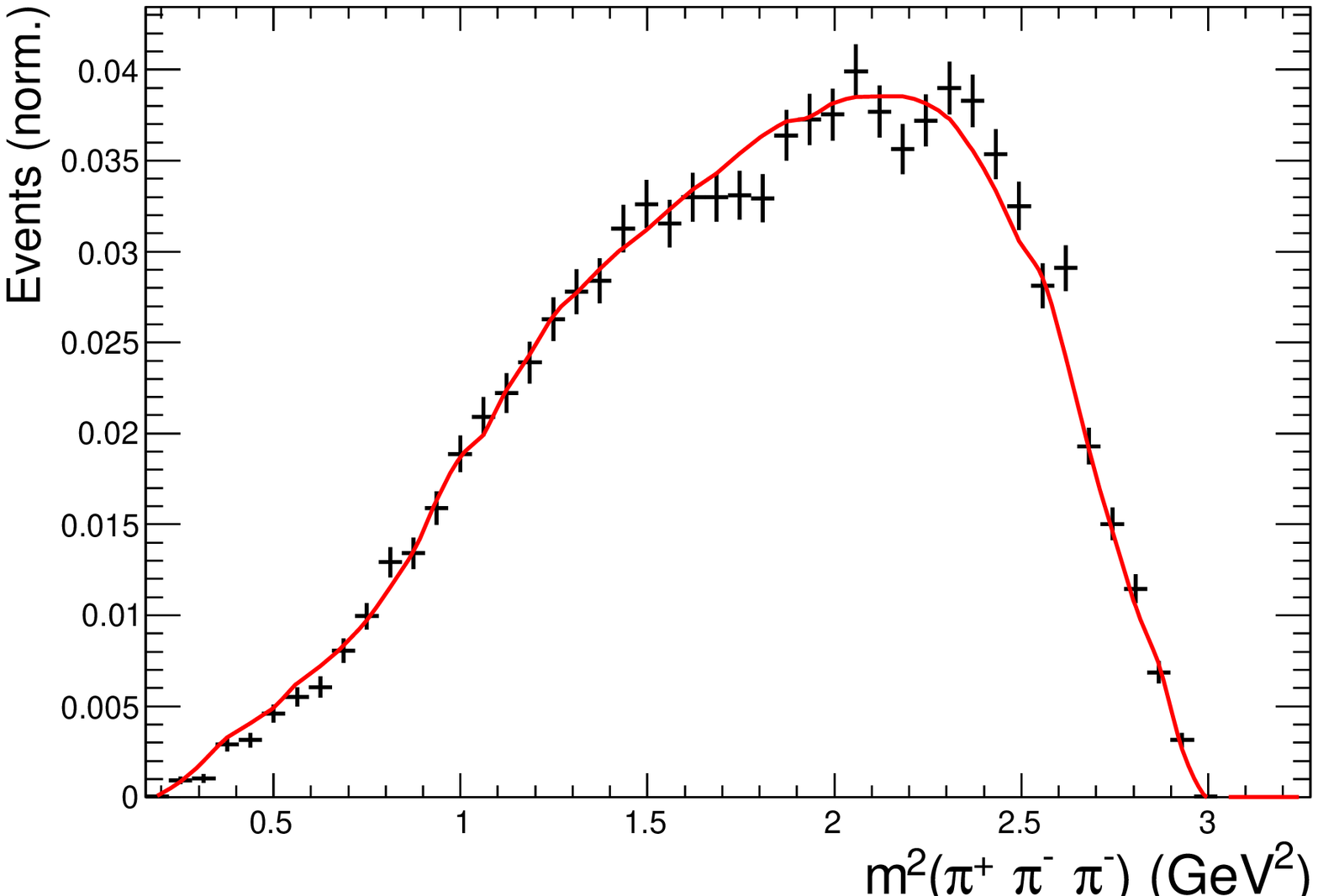} 

	\caption{Invariant mass distributions of signal events (points with error bars) and fit projections (red solid line).}
	\label{fig:baselineFit}
\end{figure}

Table \ref{tab:lassoModel} lists the real and imaginary part of the complex amplitude coefficients $a_{i}$ 
obtained by fitting the LASSO model to the data,
along with the corresponding fit fractions.
The latter are defined as 
\begin{equation}
\label{eq:DefineFitFractions}
	F_{j} = \frac{\int \left\vert   a_{j} \, \mathcal A_{j} \right\vert^{2} \,\text{d}\Phi_{4} }
	{\int \left\vert  \mathcal A_{D^{0}} \right\vert^{2} \, \text{d}\Phi_{4}} .
\end{equation}
These fractions do not have to sum to $100 \%$ due to interference.
The dominant contribution is the $a_{1}(1260)$ resonance in the decay modes
$a_{1}(1260) \to \rho(770) \pi$ and $a_{1}(1260) \to \sigma \pi$
followed by the quasi-two-body decays $D \to \sigma f_{0}(1370)$ and 
$D \to \rho(770) \rho(770)$.
Figure \ref{fig:baselineFit} shows the distributions of the phase-space observables
for data superimposed by the fit projections. 
A reasonable agreement is observed for each distribution.
In order to quantify the quality of the fit including the correlation of the phase-space variables,
a \chisq value is determined by binning the data in phase-space as
\begin{equation}
	\chi^{2} = \sum_{b=1}^{N_{\rm bins}} \frac{[N_{b}-N_{b}^{\rm exp}]^{2}}{N_{b}^{\rm exp}},
\end{equation}
where $N_{b}$ is the number of data events in a given bin, 
$N_{b}^{\rm exp}$ is the event count predicted by the fitted PDF
and $N_{\rm bins}$ is the number of bins.
An adaptive binning 
is used to ensure sufficient statistics in each bin for a robust $\chi^{2}$ calculation \cite{KKpipi}.
The \chisq value divided by the number of degrees of freedom
amounts to $\chisq/\nu = 1.33$,
indicating a good fit quality.
  
\begin{figure}[ht]
	\centering
	\includegraphics[width=0.32\textwidth, height = 4cm]{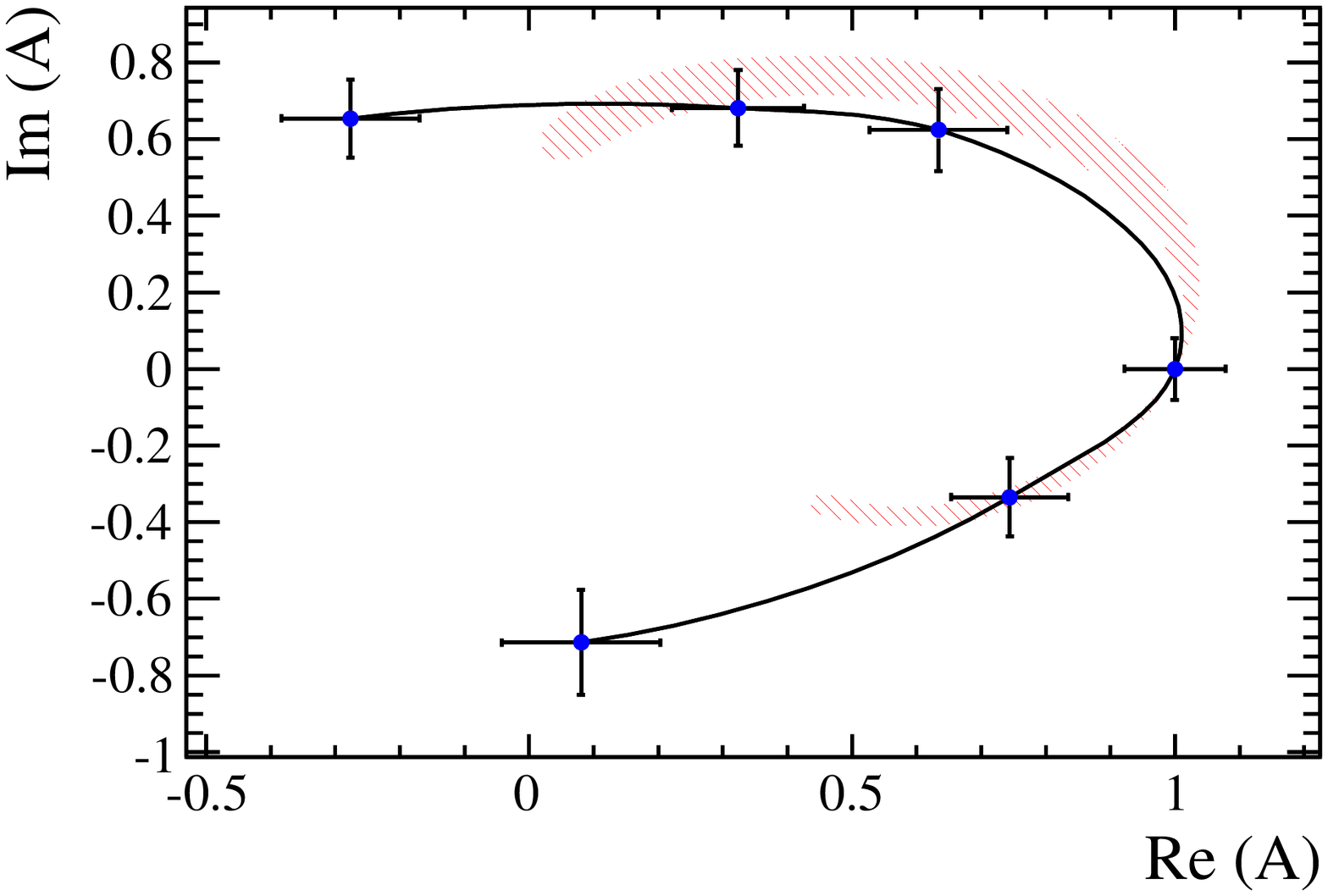} 
	\includegraphics[width=0.32\textwidth, height = 4cm]{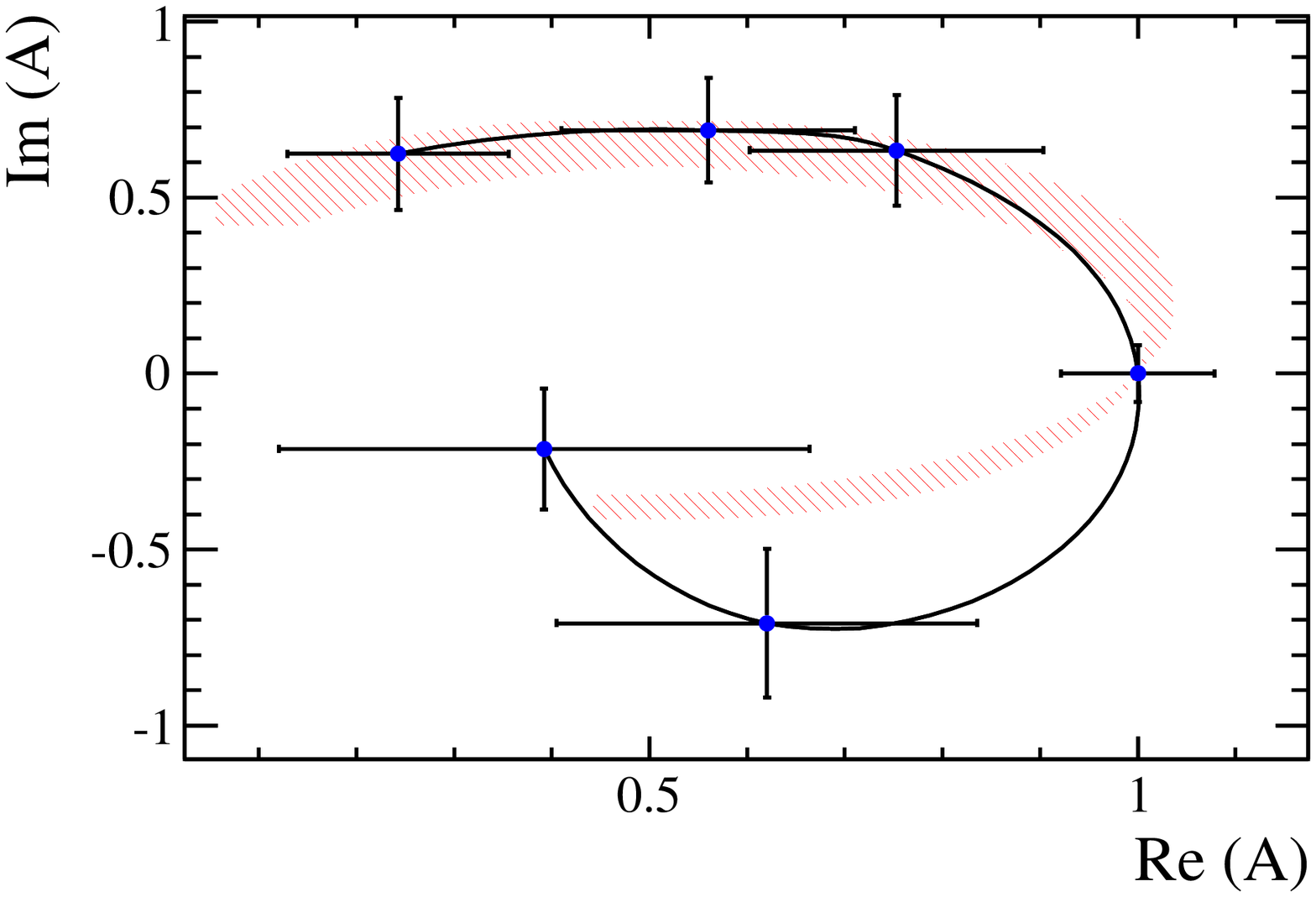} 
	\includegraphics[width=0.32\textwidth, height = 4cm]{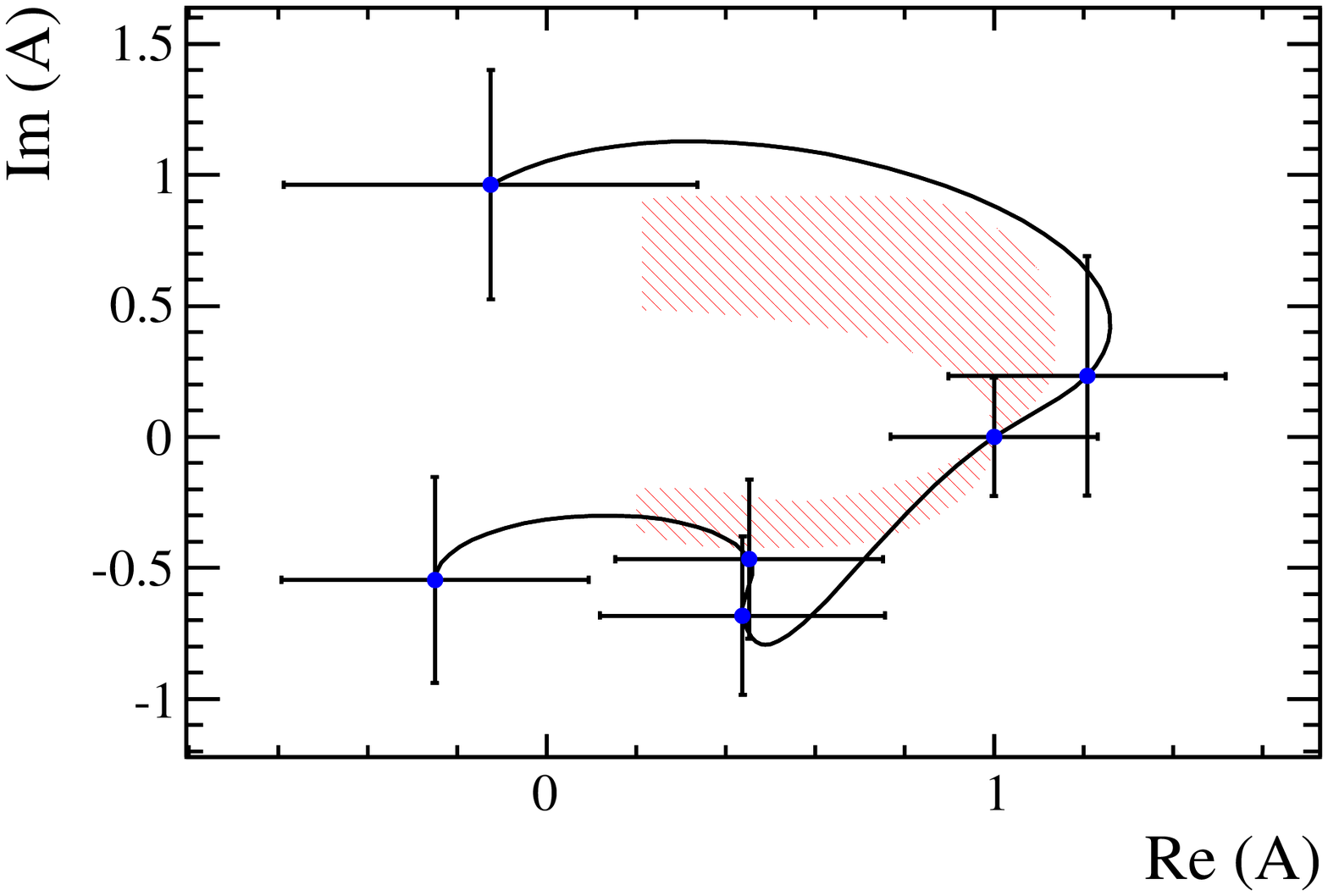} 
	\caption{\small Argand diagrams of the quasi-model-independent $a_{1}(1260)$ (left), $\pi(1300)$ (middle) and $a_{1}(1640)$ (right) line shapes. 
	In each case, the fitted knots are displayed as points with error bars and the black line shows the interpolated spline.
	The Breit-Wigner line shape with the mass and width from the nominal fit is superimposed (red area). 
	The latter is chosen to agree with the interpolated spline at the point ${\rm Re}(A) = 1$, ${\rm Im}(A) = 0$.}
	\label{fig:argand}
\end{figure}

As a cross-check, we verify 
the resonant phase motion of the observed $a_{1}(1260)$, $\pi(1300)$ and $a_{1}(1640)$ resonances in a quasi-model-independent way as pioneered in Ref.~\cite{Aaij:2014jqa}.
For this purpose, the corresponding Breit-Wigner line shapes of the resonances are replaced, one at a time, by a complex-valued cubic spline. 
The interpolated cubic spline has to pass through 
six independent complex knots
spaced in the $m^{2}(\pip \pip \pim)$ region around the nominal resonance mass.
The fitted real and imaginary parts of the knots are shown in Fig.~\ref{fig:argand},
where the expectations from a Breit-Wigner shape with 
the mass and width from the nominal fit 
are superimposed 
taking only the uncertainties on the 
mass and width into account. 
In each case, the Argand diagram 
shows a clear circular, counter-clockwise trajectory which is the expected behavior of a resonance. 
Since the investigated resonances are all very broad, the model independent line shapes 
can absorb statistical fluctuations in the data, especially near the phase-space boundaries. 
Therefore, the agreement with the Breit-Wigner expectation can be considered as qualitatively reasonable in all cases
indicating that these resonances are indeed real features of the data. 
Finally, the fractional $CP$-even content,
\begin{equation}
	F_{+}^{4\pi} = \frac{\int \vert \mathcal A_{D^{0}} + \mathcal A_{\overline{D^{0}}} \vert^{2}  \, \text{d}\Phi_{4}   }
	{\int \vert \mathcal A_{D^{0}} + \mathcal A_{\overline{D^{0}}} \vert^{2}  \, \text{d}\Phi_{4}  + \int \vert \mathcal A_{D^{0}} - \mathcal A_{\overline{D^{0}}} \vert^{2}  \, \text{d}\Phi_{4} } ,
\end{equation}
is calculated from the LASSO model to be  
\begin{equation}
F_{+}^{4\pi} (\mathrm{flavour-tagged, model-dependent}) = (73.5 \pm 0.9 \stat) \%,
\end{equation}
in excellent agreement with a previous model-independent analysis of CP-tagged events \cite{Malde:2015mha},
\begin{equation}
F_{+}^{4\pi} (\mathrm{CP-tagged, model-independent}) = (73.7 \pm 2.8) \% .
\end{equation}

\begin{table}[hp]
	\centering
	\footnotesize
	\begin{tabular}{lrrr}
	\hline
	\textbf{Decay mode} & \textbf{Re $a_{i}$} & \textbf{Im $a_{i}$} & \textbf{$F_{i} (\%)$} \\
	$D^{0} \to \pim \, \left[ a_{1}(1260)^{+}\to \pip \, \rho(770) \right] $ & 100.0 (fixed) & 0.0 (fixed) & $36.7 \pm 2.4$   \\
	$D^{0} \to \pim \, \left[ a_{1}(1260)^{+} \to \pip \, \sigma \right] $ & $43.8 \pm 4.5$ &$ 35.5 \pm 4.2$ &  $10.9 \pm 1.5$\\
	$D^{0} \to \pip \, \left[ a_{1}(1260)^{-}\to \pim \, \rho(770) \right] $ & $31.9 \pm 3.7$ & $10.7 \pm  2.8$& $4.1 \pm 0.5$   \\
	$D^{0} \to \pip \, \left[ a_{1}(1260)^{-} \to \pim \, \sigma \right] $ & $10.2 \pm 2.3$  & $16.2 \pm 2.1$  &  $1.2 \pm 0.2 $\\
	
	$D^{0} \to \pim \, \left[ \pi(1300)^{+}\to \pip \, (\pip \, \pim)_{P} \right] $ &$-17.2 \pm  2.7$&$ -37.3 \pm 5.0 $ & $6.1 \pm 0.7 $   \\
	$D^{0} \to \pim \, \left[ \pi(1300)^{+} \to \pip \, \sigma \right] $ &$ -33.4 \pm 4.4 $&$ 5.6 \pm  3.5$ &  $4.2 \pm 1.0$\\
	$D^{0} \to \pip \, \left[ \pi(1300)^{-}\to \pim \, (\pip \, \pim)_{P} \right] $ &$-25.4 \pm 4.4$ &$2.8 \pm 4.6 $ & $2.3 \pm 0.5$   \\
	$D^{0} \to \pip \, \left[ \pi(1300)^{-} \to \pim \, \sigma \right] $ & $-3.2 \pm 4.3$ & $20.6 \pm 3.4$ &  $1.6 \pm 0.4 $\\

	$D^{0} \to \pim \, \left[ a_{1}(1640)^{+}[D] \to \pip \, \rho(770) \right] $ & $-16.2 \pm 4.5$ &$28.1 \pm 8.9$&  $3.6 \pm 0.6 $\\
	$D^{0} \to \pim \, \left[ a_{1}(1640)^{+}\to \pip \, \sigma \right] $ & $0.1 \pm  0.4$&$-18.3 \pm 5.1$ & $1.2 \pm 0.5 $  \\

	$D^{0} \to \pim \, \left[ \pi_{2}(1670)^{+}\to \pip \, f_{2}(1270) \right] $  &$0.2 \pm  2.6$& $21.0 \pm 2.7$ & $1.5 \pm 0.3 $  \\
	$D^{0} \to \pim \, \left[ \pi_{2}(1670)^{+} \to \pip \, \sigma \right] $  & $-15.0 \pm 2.7$ & $-27.1 \pm 3.5$ & $3.3 \pm 0.6 $ \\

	$D^{0} \to \sigma \, f_{0}(1370)  $ &$28.3 \pm  3.4$& $69.8 \pm 5.9$&  $18.4 \pm 1.4 $ \\
	
	$D^{0} \to \sigma \,  \rho(770)  $ &$34.8 \pm 4.4 $& $-9.5 \pm 4.0$ & $4.4 \pm 1.0$\\

	$D^{0} \to \rho(770) \, \rho(770)$  & $1.0 \pm 3.0$ & $15.1 \pm 3.7 $&  $0.9 \pm 0.3 $ \\
	$D^{0}[P] \to \rho(770) \, \rho(770)$  & $-4.1 \pm 2.7$ & $- 41.6 \pm  2.6$ &   $7.1 \pm 0.5$\\
	$D^{0}[D] \to \rho(770) \, \rho(770)$ &$-66.4 \pm  5.1$&$ 0.1 \pm 3.1 $&  $15.5 \pm 1.2 $\\
	
	$D^{0} \to f_{2}(1270) \,  f_{2}(1270) $  &$-7.9 \pm  2.5$& $-15.4 \pm  2.3$&   $1.1 \pm 0.3$\\
	
	\hline
	Sum & & &  $123.7 \pm 6.8$ \\
	$m_{a_{1}(1260)} (\mev)$ & & &  $1231 \pm 8 $\\
	$\Gamma_{a_{1}(1260)} (\mev)$ & & &  $459 \pm 18 $\\	
	$m_{\pi(1300)} (\mev)$ & & &  $1180 \pm 12 $\\
	$\Gamma_{\pi(1300)} (\mev)$ & & &  $297 \pm 20 $\\
	$m_{a_{1}(1640)} (\mev)$ & & &  $1644 \pm 16 $\\
	$\Gamma_{a_{1}(1640)} (\mev)$ & & &  $222 \pm 56 $\\
	\midrule
	$\chi^{2}/\nu$ & & & $1.33$  \\
	$F_{+}^{4\pi}(\%)$ & & &  $73.5 \pm 0.9 $ \\  
	\hline
	\end{tabular}
	\caption{\small Real and imaginary part of the complex amplitude coefficients 
	and fractional contribution of each component of the LASSO model. 
	The individual amplitudes are renormalized prior to the amplitude fit such that 
        $\int  \left\vert  \mathcal A_{i} \right\vert^{2} \, \text{d}\Phi_{4} = 1$.
         The quoted uncertainties are statistical only.}
	\label{tab:lassoModel}
\end{table}

\section{Search for direct CP violation}
\label{sec:CPV}

A search for \CP violation is performed by fitting the baseline model
to the flavour tagged $D^{0}$ and $\overline{D^{0}}$ samples.
In contrast to our default fit described in \secref{sec:fit},
we now allow the amplitude coefficients for 
$D^{0} \to \fourpi$ and $\overline{D^{0}} \to \fourpi$ decays to differ.
The fit to the $D^{0}$ and $\overline{D^{0}}$ samples has to be performed simultaneously 
in order to correctly account for mistagged events.
Table \ref{tab:CPV} compares 
the resulting fit fractions for the $D^{0}$ and $\overline{D^{0}}$ decays.
For each amplitude the direct \CP asymmetry,
\begin{equation}
	A^{CP}_{i} =  \frac{F_{i} - \overline{F_{i}} }
	{F_{i} + \overline{F_{i}} }  ,
\end{equation}
is calculated from these fit fractions.
The sensitivity to $A^{CP}_{i}$ is at the level of $3 \%$ to $34 \%$ depending on the decay mode. 
No significant \CP violation is observed for any of the amplitudes. 

\begin{table}[ht]
	\centering
	\footnotesize
	\hspace{-1cm}
	\begin{tabular}{lrrr}
	\hline
	\textbf{Decay mode} & \textbf{$F_{i}(\%)$} & \textbf{$\overline{F}_{i}(\%)$} &  \textbf{$A^{CP}_{i} [\%]$}\\
	
	$D^{0} \to \pim \, \left[ a_{1}(1260)^{+}\to \pip \, \rho(770) \right] $ & $38.8 \pm 2.5$ & $35.2 \pm 2.4 $ & $5.0 \pm 3.2$   \\
	$D^{0} \to \pim \, \left[ a_{1}(1260)^{+} \to \pip \, \sigma \right]  $ & $10.9 \pm 1.5 $ & $9.9 \pm 1.3 $ & $5.0 \pm 3.2 $ \\

	$D^{0} \to \pip \, \left[ a_{1}(1260)^{-}\to \pim \, \rho(770) \right]   $ & $4.4 \pm 0.8$  & $3.9 \pm 0.8 $ &  $6.8 \pm 13.2 $ \\
	$D^{0} \to \pip \, \left[ a_{1}(1260)^{-} \to \pim \, \sigma \right]  $ & $1.2 \pm 0.3$  & $1.1 \pm 0.3 $ &  $6.8 \pm 13.2 $\\

	$D^{0} \to \pim \, \left[ \pi(1300)^{+}\to \pip \, (\pip \, \pim)_{P} \right]$ &$5.8 \pm 0.9$ & $6.8 \pm 1.0 $ &  $-7.4 \pm 8.0 $\\
	$D^{0} \to \pim \, \left[ \pi(1300)^{+} \to \pip \, \sigma \right] $ & $4.5 \pm 0.7$ & $5.2 \pm 0.8 $ &  $-7.4 \pm 8.0 $\\

	$D^{0} \to \pip \, \left[ \pi(1300)^{-}\to \pim \, (\pip \, \pim)_{P} \right]  $ & $1.9 \pm 0.5$  & $2.3 \pm 0.6$ &  $-9.6 \pm 16.5 $  \\
	$D^{0} \to \pip \, \left[ \pi(1300)^{-} \to \pim \, \sigma \right] $ & $1.5 \pm 0.4$  & $1.8 \pm 0.5 $ &  $-9.6 \pm 16.5$ \\

	$D^{0} \to \pim \, \left[ a_{1}(1640)^{+}[D] \to \pip \, \rho(770) \right] $ &$3.6 \pm 0.7$ & $3.1 \pm 0.7 $ &  $7.8 \pm 12.5 $\\
	$D^{0} \to \pim \, \left[ a_{1}(1640)^{+}\to \pip \, \sigma \right]  $ & $1.2 \pm 0.5$ & $1.0 \pm 0.5 $  &  $7.8 \pm 12.5 $\\

	$D^{0} \to \pim \, \left[ \pi_{2}(1670)^{+}\to \pip \, f_{2}(1270) \right]$ & $1.8 \pm 0.5$ & $1.6 \pm 0.5 $ &  $6.7 \pm 14.0 $\\
	$D^{0} \to \pim \, \left[ \pi_{2}(1670)^{+} \to \pip \, \sigma \right]$ & $3.3 \pm 0.8$ & $2.9 \pm 0.6 $ &  $6.7 \pm 14.0$\\

	$D^{0} \to \sigma \, f_{0}(1370)  $ &$16.9 \pm 1.6$ & $20.2 \pm 1.5 $ &  $-8.7 \pm 4.5$\\
	
	$D^{0} \to \sigma \,  \rho(770)  $ & $6.4 \pm 1.4$ & $3.8 \pm 1.1 $ & $ 26.3 \pm 15.2 $\\

	$D^{0} \to \rho(770) \, \rho(770)$  & $0.5 \pm 0.4$ & $1.3 \pm 0.6$ &   $-46.7 \pm 34.0 $\\
	$D^{0}[P] \to \rho(770) \, \rho(770)$  & $6.5 \pm 0.6$ & $7.8 \pm 0.7$ &$-9.1 \pm 7.9 $\\
	$D^{0}[D] \to \rho(770) \, \rho(770)$ & $13.9 \pm 1.7$ & $16.3 \pm 2.1 $ &   $-7.9 \pm 8.3$\\
	
	$D^{0} \to f_{2}(1270) \,  f_{2}(1270) $  & $0.9 \pm 0.4$ & $1.6 \pm 0.5$&  $-28.7 \pm 20.7$\\
	
	\hline
	\end{tabular}
	\caption{\small Fit fractions for $D^{0}$ and $\overline{D^{0}}$ decays along with the $A_{CP}$ values. Only the statistical uncertainties are given.}
	\label{tab:CPV}
\end{table}

 \clearpage


\section{Conclusions}
\label{sec:conclusions}

Preliminary results from the first amplitude analysis of flavor-tagged $D \to \fourpi$ decays 
have been presented. 
The study uses  $e^{+}e^{-}$ collision data produced at the $\psi(3770)$ resonance
corresponding to an integrated luminosity of $818 \pb^{-1}$ and 
recorded by the CLEO-c detector.
Due to the large amount of possible intermediate resonance components, a model-building procedure has been applied 
which balances the fit quality against the number of free fit parameters.
The selected amplitude model contains a total of $18$ components.
The prominent contribution is found to be the $a_{1}(1260)$ resonance in the decay modes
$a_{1}(1260) \to \rho(770) \pi$ and $a_{1}(1260) \to \sigma \pi$.
Further cascade decays involve the resonances $\pi(1300)$ and $a_{1}(1640)$.
Their line shapes have been studied in a model-independent approach and found 
to be consistent with the Breit-Wigner prediction.
The $CP$-even fraction of the decay $D \to \fourpi$ as predicted by the amplitude model 
is in excellent agreement with a previous model-independent study providing an important 
cross-check of the model.
The amplitude model has also been used to search for 
$CP$ violation in $D^{0} \to \fourpi$ and $\overline{D^{0}} \to \fourpi$ decays.
No $CP$ violation among the amplitudes is observed within the given precision of a few percent.

\section*{Acknowledgements}
This  analysis  was  performed  using  CLEO-c  data.   The  authors  of  these proceedings (some of whom were members of CLEO) are grateful to the collaboration for the privilege of using these data. We also gratefully acknowledge the support of the UK Science and Technology Facilities Council, the European Research Council 7 / ERC Grant Agreement number 307737 and the German Federal Ministry of Education and Research (BMBF).

\bibliographystyle{LHCb}
\bibliography{main}

\end{document}